\documentclass{article}
\usepackage{spconf,amsmath,graphicx}
\usepackage{cite}
\usepackage{hyperref}
\usepackage{amssymb,amsfonts}
\usepackage{mathrsfs}
\usepackage{algorithmic}
\usepackage{textcomp}
\usepackage{xcolor}
\usepackage{bm}
\usepackage{bbm}
\usepackage{caption}
\usepackage{array,multirow, makecell}
\graphicspath{{./images/}}

\title{Limitations of weak labels for embedding and tagging}
\name{Nicolas Turpault \thanks{This work was made with the support of the French National Research
	Agency, in the framework of the project LEAUDS ``Learning to understand
	audio scenes'' (ANR-18-CE23-0020) and the French region Grand-Est.
	Experiments presented in this paper were carried out using the Grid'5000
	testbed, supported by a scientific interest group hosted by Inria and
	including CNRS, RENATER and several Universities as well as other
	organizations (see \url{https://www.grid5000.fr}).} \qquad Romain Serizel \qquad Emmanuel Vincent}
\address{Universit\'e de Lorraine, CNRS, Inria, Loria, F-54000 Nancy, France\\}
\begin{document}
\ninept
\maketitle
\begin{abstract}
Many datasets and approaches in ambient sound analysis use weakly labeled data.
Weak labels are employed because annotating every data sample with a strong label is too expensive.
Yet, their impact on the performance in comparison to strong labels remains unclear.
Indeed, weak labels must often be dealt with at the same time as other challenges, namely multiple labels per sample, unbalanced classes and/or overlapping events.
In this paper, we formulate a supervised learning problem which involves weak labels.
We create a dataset that focuses on the difference between strong and weak labels as opposed to other challenges. We investigate the impact of weak labels when training an embedding or an end-to-end classifier.
Different experimental scenarios are discussed to provide insights into which applications are most sensitive to weakly labeled data.
\end{abstract}

\section{Introduction}
\label{sec:intro}
Sound carries a lot of information that can provide important information on our environment.
In recent years, interest in ambient sound analysis has grown in particular due to the numerous potential applications \cite{virtanen_computational_2017}.
Most current approaches rely on training ``big'' classifiers in an end-to-end fashion on large-scale labeled data. 
An alternative approach is to learn an intermediate representation or embedding of the data that can allow for better generalization, shorter training time and smaller labeled data requirements by separating the time-consuming stage of learning the embedding from the final stage of training a smaller classifier.
This approach has been used successfully in various domains related to audio signal processing  \cite{pascual_learning_2019,van_der_maaten_stochastic_2012}.

Several attempts towards learning meaningful embeddings have been made in the field of ambient sound analysis relying on audio-only \cite{xu_unsupervised_2017,cartwright_tricycle:_2019,pons_training_2019} or audiovisual data \cite{cramer_look_2019}.
These approaches are either unsupervised or based on a very small amount of labeled data.
Some supervised approaches have been proposed that can exploit labeled data to learn an embedding with a classifier that is later truncated \cite{hershey_cnn_2017} or to learn the embedding using sampling methods like triplet networks \cite{weinberger_distance_2009} or prototypical networks \cite{snell_prototypical_2017}.
Tokozume et al.\ \cite{tokozume_learning_2018} explained how embeddings can be learned by mixing two examples and predicting the ratio of the mix. However the extension of these approaches to weakly labeled data remains an open issue.

Weak labels consist of indicating the presence of a label in a segment without any information about the number of instances or their time localization in the recording.  This can be considered as introducing noise in the labels as opposed to strong labels that can be considered as clean, accurate labels. Weak labels form a recurring challenge in various machine learning applications \cite{lu_learning_2017,schluter_learning_2016} including ambient sound analysis \cite{kumar_audio_2016}. Since obtaining enough strongly labeled data is usually too expensive, a common choice is to gather a sufficient amount of weakly labeled data. For instance, Task 4 of the Detection and Classification of Acoustic Scene and Events (DCASE) challenge\footnote{\url{http://dcase.community/challenge2019/task-sound-event-detection-in-domestic-environments}} has provided weakly labeled data since 2018.

The use of weakly labeled data to train a sound event detection system, which outputs labels together with their time localization in a segment of audio, has been studied in recent years \cite{mcfee_adaptive_2018,serizel_sound_2019,shah_closer_2018}. A common approach is to use multi-instance learning \cite{kumar_audio_2016} to predict strong labels while training on weak labels only.
%This relies on the aggregation of predicted strong labels that gives a weak label we can compare with the groundtruth.
%In these method the loss is on weak labels with a constraint of being coherent on a strong label perspective, but this does not create strong labeled embeddings.
The top performing systems for DCASE Task 4 used a mean-teacher model \cite{lin_guided_2019,delphin-poulat_mean_2019}. However, from the evaluation reports, it is hard to analyze how much weakly labeled data and the strongly labeled data is used for training and the impact of this distribution on the performance.

Shah et al.\ \cite{shah_closer_2018} analyzed the impact of weak labels using Audioset \cite{gemmeke_audio_2017} (which is already weakly labeled) and measured the performance degradation when extending the length of the original 10~s segments to 30~s or 60~s.
Audioset has the advantage of being real data, but this has the inherent drawback of posing several additional challenges that are difficult to analyze separately: multiple labels per segment, unbalanced classes, and overlapping events.
%Tokozume et al.\ \cite{tokozume_learning_2018} exploited strongly labeled data to successfully learn an embedding.
Tokozume et al.\ \cite{tokozume_learning_2018} used data from UrbanSound8k \cite{salamon_dataset_2014}, a synthetic dataset composed of urban sound recordings extracted from Freesound \cite{font_freesound_2013} and verified by human annotators. The segments are mostly strongly labeled and last up to 4~s.

In order to properly tackle the challenge of weak labels, we need strongly labeled events in longer recordings so that we can simulate weaker labels in a controlled way.
Also, we want a single event per recording so as to focus on the challenge of weak labels and avoid multiple labels, unbalanced classes, and overlapping events which could be considered in future experiments.
To do so, we propose to create a synthetic dataset by combining ambient sound events from Freesound with a background sound. This approach is directly inspired by the DESED synthetic dataset \cite{turpault_sound_2019}.

Our contributions in this paper are the formulation of the problem specific to supervised learning using weak labels for embedding learning and tagging, and the experimental analysis using several embedding learning methods in order not to depend on a specific method. The dataset and the code are available \footnote{\url{https://github.com/turpaultn/walle}}.

The remainder of this manuscript is organized as follows. Section 2 describes the different methods to learn embeddings and perform tagging. Section 3 introduces the proposed dataset. Section 4 describes the experiments. Section 5 discusses the results and conclusions are provided in Section 6.
%learning embeddings by events, by segments of 200~ms, 1~s and 10~s and show the impact of the different classes and events.

%\begin{itemize}
%	\item Sound has rich information: pitch, magnitude, phase, harmonics, what if we could get a simple and rich representation of audio sound ? cite people doing that, for speaker identification, and speech
%	Networks learning a representation (autoencoders) or using sampling (siamese, triplets)
%
%	\item Audio Tagging is the final goal (datasets, previous methods, ..)
%	\item Speak about Desed dataset, which is weakly labeled data but with real data
%	\item Weakly supervised learning (different approaches)
%	\item Here we try to get a good representation so the recognition is easier, and show the limitation to finish on the possible improvement
%\end{itemize}
\section{Learning embeddings}
\label{sec:pb}
Let $\mathscr{C}$ be a set of $K$ classes.
We have a dataset $\mathscr{D} = \{(\bm{x}_i, \bm{y}_i)\}_{i=1}^N$ where $\bm{x}_i$ is a time-frequency representation of the input data and $\bm{y}_i= [y_{i,1}, ..., y_{i,K}]$ is a vector containing the labels with $y_{i,k} \in \{0,1\}$ indicating whether the sound event class $k$ is present in the clip or not.
Our goal is to learn an embedding $E$ that can easily discriminate the classes $k \in \mathscr{C}$.
The embedding network is followed by a classifier $G$ which performs audio tagging, i.e., detecting the sound event classes that are present within an audio clip, regardless of their time boundaries.
$E$ and $G$ can be trained jointly (end-to-end classifier) or $E$ can be trained separately from $G$ (triplet network or prototypical network). We detail these three approaches below.

\subsection{End-to-end classifier}
The end-to-end approach consists of jointly training $E$ and $G$ to minimize a classification cost. 
In the following, we minimize the binary cross-entropy
\begin{equation}\label{xent}
\sum_k -y_{i,k} \log(G(E( \bm{x}_i))_k)-(1- y_{i,k})\log(1-G(E(\bm{x}_i))_k).
\end{equation}
Since $E$ is not trained to optimize an explicit distance, there is no such distance between the embeddings learned.

\subsection{Triplet network}
In the triplet network based approach, we consider triplets $(\bm{x}^\text{a}, \bm{x}^\text{p}, \bm{x}^\text{n})$ where the anchor $\bm{x}^\text{a}$ is any sample from the training dataset, the positive example $\bm{x}^\text{p}$ is a random sample with the same label as the anchor and the negative example $\bm{x}^\text{n}$ is a random sample with a label different from that of the anchor. The embedding network $E$ is trained by minimizing the triplet loss~\cite{wang_learning_2014}
\begin{align}
\sum_{\bm{x}_i\in \mathscr{D}} \big[||E(\bm{x}_i^\text{a}) - E(\bm{x}_i^\text{p})||_2^2 - ||E(\bm{x}_i^\text{a}) - E(\bm{x}_i^\text{n})||_2^2  + \delta \big]_{+},
\end{align}
where $[.]_{+}$ is the hinge loss, $||.||_2$ is the $L_2$ norm, and $\delta$ is a margin parameter. The triplet loss aims to find a meaningful embedding space in which the anchor and the positive example are closer than the anchor and the negative example. The margin corresponds to the difference of the distance between the anchor and the negative example and the distance between the anchor and the positive example (in the embedding space). The larger the margin, the further the negative example will be. Since the margin depends on distances between the embeddings, in order for it to make sense, the embeddings must be normalized before computing the distances.

\subsection{Prototypical network}
In the prototypical network based approach, the data are sampled in a specific manner. Each batch contains $J$ classes (in the following, $J=K$), and $m$ training data points $(\bm{x}_i, \bm{y}_i)$ for each class. Among them, $m_s$ points are called support points and are used to generate a \textit{prototype} of the class. The prototype vector for class $j$ is the average of the embeddings of the support points:
\begin{equation}
	\label{proto}
	\bm{c}_j = \frac{1}{m_j}\sum_{(\bm{x}_i, \bm{y}_i)\in \mathscr{D}_j} E(\bm{x}_i)
\end{equation}
where $\mathscr{D}_j$ is the set of support points of that class.
The remaining $m_q = m - m_s$ points are called query points and are used to train the embedding network $E$. The loss function to be minimized is the sum over all queries of the cross-entropy between the class label of a given query and the softmax over the distances between the embedding of that query and the prototypes of all classes. This results in a soft assignment of each query point to one of the $J$ classes based on these distances.

\subsection{Classification of the embeddings}
Once the embeddings have been learned with the triplet network or the prototypical network, we can learn a classifier $G$ from these embeddings by optimizing the cross-netropy cost in~\eqref{xent}.

\section{Datasets}
\label{sec:dataset}
In order to analyze the impact of weak labels independently of other challenges, we introduce two new datasets called the weak annotation analysis (WAA) dataset and the 200~ms dataset. These datasets are generated from the DESED dataset \cite{turpault_sound_2019} and contain 10~s sound clips generated by mixing foreground sound events from Freesound \cite{Fonseca2017freesound,font2013freesound} with backgrounds from SINS \cite{Dekkers2017} and MUSAN \cite{snyder2015musan}. 
The synthetic data for training and validation use the foreground sound events and backgrounds from the DESED synthetic development dataset. The sound clips in our evaluation set use the foreground sound events and backgrounds from the synthetic evaluation set in DESED.
Both the training set, the validation set and the evaluation set were created using Scaper \cite{salamon2017scaper}.
We further made sure that the foreground sound events do not overlap between the training set and the validation set and that foreground files from the same Freesound user do belong to the same set. 
Since we want to focus on the problem of weak labels, we created sound clips with a single event and a signal-to-noise ratio (SNR) between the foreground sound event and the background uniformly drawn between 6~dB and 30~dB.

\begin{table}[t!]
	\centering
	\begin{tabular}{l||c|c||c}
		\multirow{ 2}{*}{Class}& \multicolumn{2}{c||}{Dev}& \\
	   &Train&Valid&Eval\\
		\hline\hline
		Alarm/bell/ringing & 177 & 13 & 63\\
		Blender 				& 89 & 9 & 27\\
		Cat 					& 78 & 10 & 26\\
		Dishes 					& 99 & 10 & 34\\
		Dog 					& 121 & 15 & 43\\
		Electric shaver/toothbrush & 51 & 5 &17\\
		Frying 					& 56 & 8 &	17\\
		Running water 	& 59 & 9 & 20\\
		Speech 				& 117 & 11 & 47\\
		Vacuum cleaner & 62 & 10 &	20\\
		\hline\hline
		Total & 909 & 100 & 314\\
	\end{tabular}
	\caption{Unique Freesound sound events used in each set.}
	\label{tab:uniquenb_sets}
\end{table}

\subsection{WAA dataset}
The WAA dataset contains 2,700 clips for training, 300 for validation and 750 for evaluation.
It is composed of the 10 sound event classes of the DESED dataset. The number of unique Freesound sound events used to generate each subset is presented in Table \ref{tab:uniquenb_sets}. Each of the subsets is balanced, i.e., it contains the same number of clips for each class.

\begin{figure}[htb]

	\begin{minipage}[b]{1.0\linewidth}
		\centering
		\centerline{\includegraphics[width=8.5cm,trim=8 8 8 30,clip]{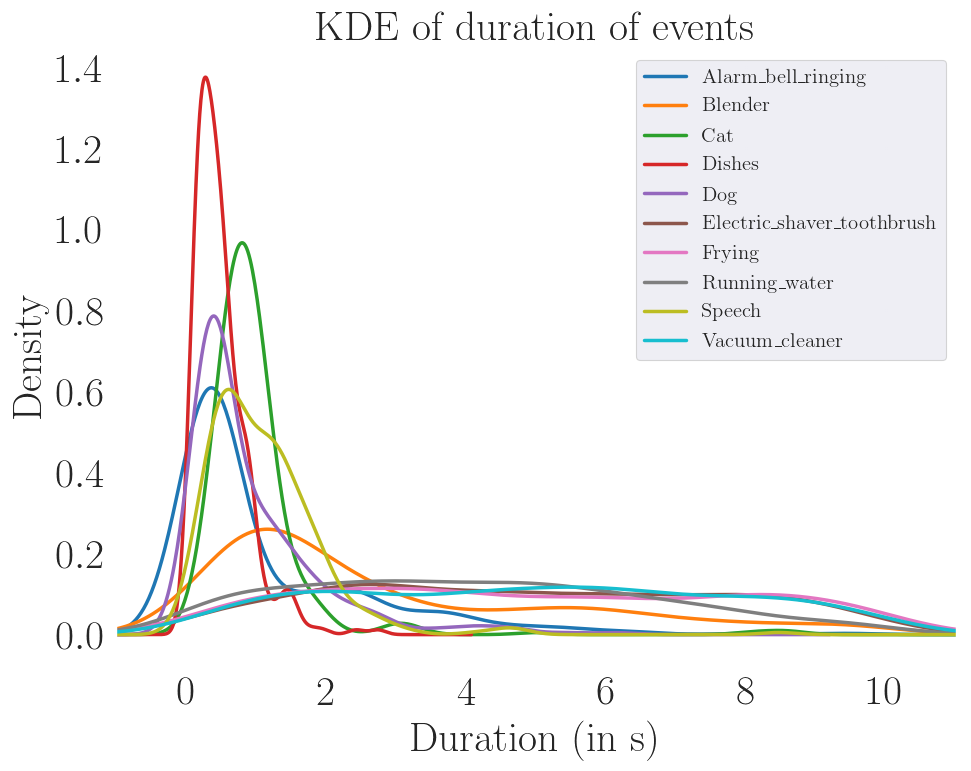}}
		%  \vspace{2.0cm}
	\end{minipage}
	\caption{Kernel density estimates of the duration of sound events for each of the 10 classes in the training dataset.}
	\label{fig:duration}
\end{figure}

The effective duration of each sound event depends on the duration of the isolated event and its onset time within the clip (sound events which are longer than the time remaining until the end of the clip are cut down).
Figure \ref{fig:duration} shows the distribution of sound event durations for each class in the training set. We can distinguish two categories of sound event classes: short events (Dishes, Speech, Alarm/bell/ringing, Dog, Cat) and long events (Blender, Electric shaver/toothbrush, Frying, Vacuum cleaner, Running water). However, the duration still varies within each category.

% \subsubsection{WAA1: ground truth segmentation}
%n this scenario, we assume that we already know the sound event onset and offset and keep only the frames where the sound event actually does occur. We learn an embedding on these frames.
%The duration of the sound clip is the same as that of the sound event so it can vary from one sound clip to the other. In this scenario, the labels can be considered as strong labels.

%\subsubsection{WAA1: fixed sized segments}
%In order to mitigate the impact of duration variability on the results, we propose to consider a fixed duration dataset which we call WAA1.
In order to control the ``weakness'' of the labels, we cut the 10~s clips into shorter fixed-sized segments.
Specifically, we assume that we know the labels, and divided the sound clips into segments that contain either the full sound event or part of it when the event is long. In order to keep a consistent subset size, we kept only one segment per original 10~s clip.
When the segment duration is small, most of the frames within the segment do actually contain the sound event therefore the label can be considered as strong. When the segment duration is larger, the number of frames where the sound event is absent increases therefore the label becomes weaker.

We set the segment durations to 200~ms, 1~s, and 10~s in practice.
Since most of the events are longer than 200~ms, the 200~ms subset is almost entirely strongly labeled.
The experiments using 1~s segments introduce some weak labels for short sound events. When the segment duration is increased to 10~s the proportion of labels that can be considered as weak increases too.

\subsection{200~ms dataset}
In order to study the impact of the weak labels on the embedding quality regardless of the event duration, we derived an additional set from the original isolated sound events which we call the 200~ms dataset.
To create this dataset, we used the same association of foreground sound events and background files as in the WAA dataset. However, we cut the duration of the foreground sound events down to 200~ms.
In this dataset, only 122 clips have a foreground event that lasts less than 200~ms. The distribution of sound events that are shorter than 200~ms is as follows: 60 Dishes, 18 Speech, 18 Dog, 13 Alarm\_bell\_ringing, 10 Running\_water, and 3 Cat events.
Therefore, even though the bias induced by sound event duration is greatly reduced, there are still about $\tfrac{1}{4}$ of the Dishes events for which the models may exhibit a different behavior.

\section{Experiments}
\label{sec:expe}
 \subsection{Feature extraction}
 The sound clips are single-channel and sampled at 44.1~kHz. We first resample them at 16~kHz.
 We then compute the short-time Fourier transform on 25~ms windows with a step size of 10~ms.
 We finally compute log-mel features with 64 mel bands.

\subsection{Model and parameters}
Our embedding model $E$ is a convolutional neural network (CNN) with 4 layers. We apply the CNN to 200~ms intervals and we average the outputs along the time axis to obtain a single embedding vector.
When considering 1~s or 10~s input segments, the embeddings obtained over each 200~ms interval are averaged over time to obtain a single embedding representing the whole segment.
We consider averaging for aggregation here as we want the approach to be extensible to the multi-event case in the future (which is not compatible with, e.g., maximum-based aggregation).
The final output is a vector of dimension 130 regardless of the duration of the input segment.

The classifier $G$ is a fully connected layer of size 32 with leaky ReLU activation followed by an output layer of size 10 with sigmoid activation which predicts the sound event class. The sigmoid is used here in order to allow for a future extension to multi-label classification.
The architecture of the model (number of convolution layers, dropout, embedding size, number of fully connected layers) has been optimized on the end-to-end classifier with the Asynchronous Successive Halving Algorithm (the asynchronous version of Hyperband)\cite{li_massively_2018}, using Orion\footnote{\url{https://github.com/Epistimio/orion}}.
We use the same architectures for all experiments.

\subsection{Validation of the embeddings for early stopping}
When the embeddings are learned separately, we perform early stopping based on a metric related to the quality of the embeddings on the validation set. We propose a metric that relies on the average value of the embeddings for each sound event class:
\begin{equation}
	\bm{c}_k = \frac{1}{|\mathscr{D}_k|}\sum_{(\bm{x}_i, \bm{y}_i)\in \mathscr{D}_k} E(\bm{x}_k)
\end{equation}
where $\mathscr{D}_k$ is the subset of $\mathscr{D}$ that contains the points $(\bm{x}_i, \bm{y}_i)$ from the class $k$ and $|\mathscr{D}_k|$ denotes the size of that subset. Note that this is similar to the prototypes~\eqref{proto} but computed on the whole training set.

In order to measure the quality of an embedding, we then define a metric that indicates for each example $(\bm{x}_i, \bm{y}_i)$ whether the closest centroid $\bm{c}_k$ is the centroid corresponding to the sound event class that is present in the sound clip:
\begin{align}  F(\bm{x}_i) = \begin{cases}
1&\text{if } \displaystyle \arg\min_k||E(\bm{x}_i) - \bm{c}_k||_2^2 = \displaystyle \arg\max_k(\bm{y}_i) \\
0&\text{otherwise.}
\end{cases}
\end{align}
This metric is motivated by the fact that we want embeddings which are grouped into separable clusters. Indeed, if every point of each class is closer to the mean of its class, we should be able to separate the classes.

\section{Results}
We have described three different methods to learn embeddings and perform sound event tagging. In this section, we report the results achieved by the final classifier $G$ on the evaluation set in terms of the F-measure.
Note that, since we are using a sigmoid output for each class, a classifier that performs randomly will likely predict that no class is active. Therefore, the F-score can be as low as 0\%, rather than 10\% as usually expected for a 10-class classification problem (this would be the case if we were using a softmax output). Also note that we are not interested in the absolute performance of the three methods but in their behavior relatively to weakly labeled data.

\label{sec:results}
We present the results on the 200~ms dataset in Table~\ref{res:200ms}. Training the models with 200~ms segments (first row for each method), and predicting aggregated embeddings for 1~s or 10~s segments containing background noise does not work well. Therefore, if we have a strongly labeled training dataset and we train a model on a good segmentation, we cannot expect it to predict accurately the labels on unsegmented data.
Training the models on weakly labeled data also has an impact even when we test the models on already segmented data (first column of the table). This impact is a lot more negative on the embedding methods based on triplets and prototypes than on the end-to-end classifier, possibly because when using segments that are longer than the actual event we are mostly learning embeddings for the background noise. 
By contrast, training the end-to-end classifier on 1~s segments actually improves the performance when testing on 200~ms segments. This could be due to the noisy frames acting as a regularization to the model which sees a small amount of data (especially in the case of short segments).

% \begin{table}[]
% 	\setlength\extrarowheight{1ex}
% 	\begin{tabular}{llll}
% 		& 200~ms   & 1     & 10    \\ \hline
% 		Classifier & 36.3$\pm$1.7 & 45.1$\pm$2.4 & 19.0$\pm$2.1 \\ \hline
% 		Triplets   &   37.7$\pm$1.4    &   48.3$\pm$2.0    &  19.8$\pm$0.4     \\ \hline
% 		Prototypes & 29.0$\pm$1.5 & 41.3$\pm$2.4 & 15.1$\pm$2.7 \\ \hline
% 	\end{tabular}
% 	\caption{Results on the WAA dataset}
% 	\label{res:segm}
% \end{table}

\begin{table}[]
		\setlength\extrarowheight{0.5ex}
	\begin{tabular}{|l|l|lll|}
		\hline
		\multirow{2}{*}{Method}     & \multirow{2}{*}{\begin{tabular}[c]{@{}l@{}}Training \\ segment\end{tabular}} & \multicolumn{3}{l|}{Test segment} \\ \cline{3-5}
		&                                                                           & 200~ms       & 1~s       & 10~s      \\ \hline \hline
		\multirow{3}{*}{Classifier} & 200~ms                               & 45.8$\pm$2.9     & 29.6$\pm$1.7     & 3.7$\pm$0.5     \\ \cline{2-5}
		& 1~s                                                                       & 44.2$\pm$1.8    & 47.4$\pm$3.2     & 12.7$\pm$2.4     \\ \cline{2-5}
		& 10~s                                                                      & 39.8$\pm$1.9     & 49.3$\pm$3.2     & 36.7$\pm$3.8     \\ \hline \hline
		\multirow{3}{*}{Triplets}   & 200~ms                           & 42.5$\pm$1.0    & 2.6$\pm$0.4     & 0.0$\pm$0.0     \\ \cline{2-5}
		& 1~s                                                                       & 39.1$\pm$2.4    & 28.9$\pm$2.7    &0.1$\pm$0.1         \\ \cline{2-5}
		& 10~s                                                                      & 0.0$\pm$0.0        & 0.0$\pm$0.0        & 0.0$\pm$0.0         \\ \hline \hline
		\multirow{3}{*}{Prototypes} & 200~ms                            & 41.2$\pm$3.5     & 9.4$\pm$2.7      & 0.0$\pm$0.0         \\ \cline{2-5}
		& 1~s                                                                       & 38.8$\pm$1.8     & 36.1$\pm$2.1    & 1.1$\pm$1.3      \\ \cline{2-5}
		& 10~s                                                                      & 0.0$\pm$0.0        & 0.0$\pm$0.0        & 0.0$\pm$0.0       \\ \hline
	\end{tabular}
\caption{F-measure (\%) achieved on the 200~ms dataset.}
\label{res:200ms}
\end{table}

We present the results on the WAA dataset in Table \ref{res:var}. We can see that, for each of the models, training on 1~s segments and testing on 1~s segments gives the best results. We can relate this observation to Figure \ref{fig:duration}. As we can see, the duration of most of the short sounds is around 1~s so the bias introduced when training on 1~s segments remains small. This is confirmed when comparing these results with those in Table~\ref{res:200ms}. On the 200~ms dataset, changing the duration of the training segments from 200~ms to 1~s and testing on 200~ms segments degrades the performance because 9 frames out of 10 during training contain noise. On the WAA dataset, when we test on 200~ms segments, training on 1~s segments performs at least as well as training on 200~ms segments (triplets) and often performs event better (end-to-end classifier and prototypes). The latter aspect also indicates that 200~ms is probably not long enough to accurately identify the sound classes.

We can also assume that 1~s is sufficient to get enough information about long events. When we train on 10~s segments, performance with the embedding-based methods degrades severely while the end-to-end classifier still performs well. Indeed, learning embeddings on longer segments becomes very complicated probably because in most cases the segments then contain mostly noise. The embeddings learned at segment level are then probably representing the background noise more than the foreground sound event class that is hardly present within the segment. The training of the classifier on the other hand is based on a decision about the class present in the segment. Since the background noise is not a class, the classifier cannot be biased towards it and it remains more robust to loose segmentation.

\begin{table}[]
	\setlength\extrarowheight{0.5ex}
	\begin{tabular}{|l|l|lll|}
		\hline
		\multirow{2}{*}{Method}     & \multirow{2}{*}{\begin{tabular}[c]{@{}l@{}}Training \\ segment\end{tabular}} & \multicolumn{3}{l|}{Test segment} \\ \cline{3-5}
		&                                                                           & 200~ms       & 1~s       & 10~s      \\ \hline \hline
		\multirow{3}{*}{Classifier} & 200~ms                               & 45.8$\pm$2.9     & 49.0$\pm$4.1     & 26.8$\pm$3.1     \\ \cline{2-5}
		& 1~s                                                                       & 46.9$\pm$1.2     & 57.5$\pm$2.5     & 38.0$\pm$1.9     \\ \cline{2-5}
		& 10~s                                                                      & 40.2$\pm$2.0     & 54.2$\pm$0.7    & 51.0$\pm$2.3     \\ \hline \hline
		\multirow{3}{*}{Triplets}   & 200~ms                               & 42.5$\pm$1.0     &38.2$\pm$3.6     & 11.7$\pm$3.2    \\ \cline{2-5}
		& 1~s                                                                       & 41.7$\pm$7.0     & 44.8$\pm$10.9    & 18.3$\pm$7.3     \\ \cline{2-5}
		& 10~s                                                                      & 9.1$\pm$3.2      & 10.2$\pm$2.0    &2.8$\pm$0.7     \\ \hline \hline
		\multirow{3}{*}{Prototypes} & 200~ms                            & 41.2$\pm$3.5     & 36.1$\pm$7.3     & 9.5$\pm$4.3      \\ \cline{2-5}
		& 1~s                                                                       & 45.2$\pm$0.4    & 52.4$\pm$3.9     & 22.0$\pm$3.4     \\ \cline{2-5}
		& 10~s                                                                      & 29.9$\pm$6.2     & 35.8$\pm$10.9    & 28.6$\pm$11.0    \\ \hline
	\end{tabular}
\caption{F-measure (\%) achieved on the WAA dataset.}
\label{res:var}
\end{table}

% Now if we compare the results of the table \ref{res:200ms} and table \ref{res:var} we can see the impact of weak labels. Indeed, because of the duration of sound events, the Table \ref{res:var} has drastically less noise in the labels during training and during testing.
% If we take the results where the training segments were longer than the testing segments (which is usually the scenario when we have weak labels but want finer precisions), representing the bottom left of each subtables, we can see that the weak labels have usually a huge impact in comparison to the case where we test on the same size we do the training (results in the diagonal).

\section{Conclusion}
In this paper, we studied the impact of learning embeddings for audio tagging on weakly labeled data. We proposed two complementary datasets composed of synthetic sound clips. We showed that weak labels degrade the performance slightly when using an end-to-end classifier trained in a discriminative manner. Learning embeddings by sampling and comparing distances (protypical network, triplet loss) is very sensitive to the bias introduced when using weak labels (i.e., several frames within the clip actually do not contain the sound event class but just some background). We observed that learning on shorter duration segments reduces this bias.  However, the amount of information contained in the segments can then become insufficient for accurate sound tagging. We also showed that using clips that are too long (both at training and test time) is introducing too much bias and that embedding-based methods then become unreliable. This work could be extended by analyzing more in detail the impact of the sound event duration. The work so far has focused on clips with a single event but real scenarios often include several sound events that possibly overlap. The impact of these aspects will also have to be investigated.
\bibliographystyle{IEEEbib}
\bibliography{datasets,refs}

\end{document}